\documentclass[a4paper,fleqn,usenatbib]{mnras}

\usepackage{newtxtext,newtxmath}

\usepackage[T1]{fontenc}
\usepackage{ae,aecompl}

\usepackage{graphicx}	
\usepackage{amsmath}	
\usepackage{amssymb}	
\usepackage{array}
\usepackage{here}
\usepackage{mathtools}
\usepackage{float}
\usepackage{siunitx}

\DeclareSIUnit\pc{pc}
\DeclareSIUnit\mJy{mJy}
\DeclareSIUnit\yr{yr}
\DeclareSIUnit\deg{deg}
\DeclareSIUnit\erg{erg}
\newcommand{\xupdownarrow}[1]{%
  {\left\updownarrow\vbox to #1{}\right.\kern-\nulldelimiterspace}
}

\title[Imprints of quasars duty cycle on the 21cm signal]{Imprints of quasar duty cycle on the 21cm signal from the Epoch of Reionization}

\author[F. Bolgar et al.]{
Florian Bolgar$^{1}$\thanks{E-mail: florian.bolgar@obspm.fr},
Evan Eames$^{1}$\thanks{E-mail: evan.eames@obspm.fr},
Cl\'ement Hottier $^{2}$\thanks{E-mail: clement.hottier@obspm.fr},
Benoit Semelin$^{1}$\thanks{E-mail: benoit.semelin@obspm.fr}.\\
$^{1}$LERMA, Observatoire de Paris, Sorbonne Universit\'e, PSL research university, CNRS, F-75014, Paris\\
$^{2}$GEPI, Observatoire de Paris, Universit\'e PSL, CNRS, 5 Place Jules Janssen, 92190 Meudon, France
}

\date{Accepted XXX. Received YYY; in original form ZZZ}

\pubyear{2018}

\begin{document}
\label{firstpage}
\pagerange{\pageref{firstpage}--\pageref{lastpage}}
\maketitle

\begin{abstract}
Quasars contribute to the $21$-cm signal from the Epoch of Reionization (EoR) primarily through their ionizing UV and X-ray emission. However, their radio continuum and Lyman-band emission also regulates the $21$-cm signal in their direct environment, potentially leaving the imprint of their duty cycle.

We develop a model for the radio and UV luminosity functions of quasars from the EoR, and constrain it using recent observations. Our model is consistent with the $z\sim 7.5$ quasar from \citet{BanadosQuasar}, and also predicts only a few quasars suitable for 21-cm forest observations ($\sim 10$ mJy) in the sky. We exhibit a new effect on the 21-cm signal observed against the CMB: a radio-loud quasar can leave the imprint of its duty cycle on the 21-cm tomography. We apply this effect in a cosmological simulation and conclude that the effect of typical radio-loud quasars is most likely negligible in an SKA field of view. For a $1-10\text{mJy}$ quasar the effect is stronger though hardly observable at SKA resolution. Then we study the contribution of the lyman band Ly-$\alpha$ to Ly-$\beta$) emission of quasars to the Wouthuisen-Field coupling. The collective effect of quasars on the $21$-cm power spectrum is larger than the thermal noise at low $k$, though featureless. However, a distinctive pattern around the brightest quasars in an SKA field of view may be observable in the tomography, encoding the duration of their duty cycle. This pattern has a high signal-to-noise ratio for the brightest quasar in a typical SKA shallow survey.
\end{abstract}
\begin{keywords}
radiative transfer, methods: numerical, dark ages, reionization, first stars, quasars: supermassive black holes
\end{keywords}



\section{Introduction}
During the Epoch of Reionization (hereafter EoR) patches of neutral hydrogen in the intergalactic medium (IGM) emit a signal in the local $21$-cm line. This signal can 
be observed either in emission or absorption against an extended background source, usually the CMB, but also against bright point sources in the case the $21$-cm forest. The detection of this $21$-cm
signal, predicted to exhibit brightness temperature fluctuations up to a few tens of mK on scales $< 1^\circ$, is the main goal of several existing instruments (such as LOFAR, PAPER, MWA and others). Several upper limits have been established \citep{Paciga13,Dillon15,Ali15,Patil17} that already rule out some of the less likely reionization models \citep[e.g.][]{Pober15}.
The accurate measurement of the statistical properties contained in its power spectrum should be achieved with the SKA and HERA, and the former should also be able to image the signal in 3 dimensions (thanks to the cosmological redshifting of the line) with a sensitivity of the few mK on scales of a few arc-minutes. When the data are available, their interpretation will rely on two building blocks, a robust modelling of the signal and an efficient method to derive constraints in the parameter space from the observed data. 

The theory that allows us to compute the intensity of the signal emitted at a given point in space is well established (see \citet{Furlanetto06} for a comprehensive review), to the point where the approximations that went into the derivation are sometimes forgotten. Most of these are well justified (like dropping second order terms in ${v \over c}$) while other are more tenuous. Of the latter kind, we can quote the assumption that the local physical quantities do not vary within the thermal line width (untrue for mini-halos), that the optical depth of the signal is much smaller than 1 (it can reach above 0.1 for strong $21$-cm forest absorbers, see \citet{Semelin16}), or that the population levels of the ground state of hydrogen are in equilibrium (the time scale to reach equilibrium is $\sim 300\, 000$ yr, set by the intensity of the CMB). One assumption that has, to our knowledge, never been discussed, is the fact that, in the computation of the spin temperature, the CMB is the only source of local $21$-cm photons that affect the regulation of the hyperfine level populations via absorption and stimulated emission. In fact, any other source of radio continuum, such as radio-loud quasars, could contribute. Then, the question becomes the following: is neutral IGM present close enough to the quasar (i.e. is the ionised bubble small enough) so that the radio emission of the quasar dominates over the CMB, and if so, how far from the radio source does that effect extend and by how much does it modify the $21$-cm signal. To answer this question we will build a quasar population model consistent with existing high redshift observations and include its contribution in the computation of the spin temperature.

Beyond the approximations used in computing the intensity of the $21$-cm signal from the relevant local quantities, the modelling depends heavily on the assumed properties of the population of sources emitting the various radiation that regulates the intensity of the $21$-cm signal. These types of radiations are of course ionising radiation that carve nul-signal bubbles, but also Lyman band radiation responsible for the Wouthuysen-Field coupling, X-rays that are able to heat up the neutral gas, and as mentioned above, the radio continuum that may, close to the source, add to the CMB background. In current modelling techniques, the modelling of the source population is rather crude. The contribution of stars to the ionizing UV background is often computed using a uniform  and non-evolving Initial Mass Function, and a similarly fixed mass-luminosity ratio is applied to the hosting dark matter halo mass, as for example in \citet{Mellema06} and subsequent works, as well as in \citet{Mesinger11} (where the fixed value for the $\zeta$ ionizing efficiency parameter is an equivalent assumption). Even in numerical simulations where the star formation rate is computed from the local state of the gas (such as in \citet{Baek2009} and subsequent works), a poor mass resolution results from the large required volume  and the limited computing power. Thus, crucial processes such as feedback and metal enrichment are not correctly taken into account. The modelling of the production of X-rays is possibly even rougher, taking little account of the much higher variability between halos of similar mass in the case of production by quasars. The spectral properties of X-rays have been somewhat explored by considering sources of different natures such as quasars and X-ray binaries \citep{Fialkov14, 21SSD}, but the ratio of the contributions of different type of sources has been kept constant in time and space. Thus, exploring a more detailed modelling of
the source population is an interesting avenue of research. It is particularly interesting for quasars because i) they may impact the signal through all 4 of the frequency ranges mentioned above and ii) they are even more difficult to include self-consitently in hydrodynamical simulations than stellar populations as they are single objects regulated by very small scale processes.

In this work we present a simple model of the quasar population that matches current observations at $z<6$ and extrapolate it to higher redshifts. Even though we could study the impact of this modelling on the 21-cm model through all four radiations types (ionizing UV, X-rays, Lyman-band and radio), we will focus in this work on the radio and Lyman-band contributions. Indeed, the ionizing UV and X-rays contributions are essentially cumulative over the history of reionization as cooling processes in neutral regions and recombination in the diffuse ionized medium are negligible. Then, the contribution of quasars to ionizating and heating the IGM is somewhat degenerate with that of other sources.  On the contrary, the Lyman-band and radio contributions that regulate the spin temperature in the neutral IGM are essentially instantaneous contributions (return to equilibrium occurs on a $300\, 000$ yr time scale). While other types of source do contribute, quasars are characterized by their high variability due to their duty cycle. A sudden rise or drop in the local Lyman-$\alpha$ flux and radio continuum around a quasar turning on or off may leave
a distinct imprint on the $21$-cm lightcone. This is what we study in this work. In a future work, we will self-consistently include the ionizing UV and X-ray contributions as predicted from the population model. 

In section 2 we briefly present the code and the simulation used in this work. Section 3 details the quasar model and compares its predictions to observations at $z<6$. Section 4 shows how the computation of the $21$-cm signal is modified in the presence of radio-loud quasar and estimates the impact on the predicted $21$-cm signal for our quasar population model. Section 5 estimates the contribution of the quasar to the Wouthuysen-Field effect and its effect on both the tomography and power spectrum of the $21$-cm signal. Finally, we summarize our conclusion in section 6.

\section{Simulation}
\subsection{Simulation setup}
The simulation we use is performed using the LICORICE code (for a detailed description, see \cite{Baek2009}). LICORICE is a particle-based code using the Tree-SPH method to model the dynamics, which is fully coupled to the radiative transfer of ionizing UV and X-rays. The radiative transfer is computed using a Monte-Carlo scheme. The simulation that we use in this work follows the evolution of $1024^3$ particles in a  $200h^{-1}$cMpc box, half baryon and half dark matter, allowing us to resolve haloes down to a few $10^9M_\odot$. We use a standard $\Lambda$CDM cosmology with $H_0=67.8\text{km/s}^{-1}$, $\Omega_M=0.308$, $\Omega_b=0.0484$, $\Omega_\Lambda=0.692$, $\sigma_8=0.8149$ and $n_s=0.968$. This simulation is taken from the 21SSD database (see \cite{21SSD}), with an efficiency of X-ray production $f_x=1$. The chosen value of the hard-to-soft ratio $r_{\text{H/S}}$ is $0$. That is, we only include soft X-rays produced for example by quasars and no hard X-ray contribution from X-ray binaries for example. Notice that the X-ray and ionizing UV contributions of QSO rely on the same modelling as in \citet{21SSD}. The more detailed QSO luminosity function modelling detailed below is used only for the radio and Lyman-band emission, which is much more sensitive to the variability due to the duty cycle and whose contribution can be computed in post-processing. Integrating the detailed modelling for all emission wavelengths will be the subject of a future work.
The Lyman band emissivity efficiency is chosen to be $f_\alpha =0.5$. This means that, compared to the spectral energy distribution of the fiducial stellar population used in \citet{21SSD}, we use half the Lyman band emissivity to ionizing emissivity ratio. The choice of $f_\alpha=0.5$ is justified by the fact the effects studied in this work are more visible at high redshift ($z > 9$), where a top-heavy IMF with a harder spectrum than the Salpeter IMF of the fiducial population is relevant.
\subsection{The Ly-$\alpha$ transfer simulation}
The 3D transfer through the IGM of Lyman-band photons produced by stars and their scattering in the Ly-$\alpha$ to Ly-$\zeta$ lines is computed using the (currently distinct) version of LICORICE described in \cite{Semelin2007} and \cite{Vonlanthen2011}. 
Estimating the local flux at Ly-$\alpha$ frequency everywhere in the simulation box is necessary to accurately compute the spin temperature of hydrogen and thus the brightness temperature of the 21-cm signal. 
Since the kinetic heating of the neutral IGM by Ly-$\alpha$ photons is weak (\cite{Furlanetto2006}) compared to that of X-rays (for our choice of X-ray luminosity), we neglect the feedback of the Ly-$\alpha$ flux on quantities other than the spin temperature and run the Lyman lines transfer in post-treatment.\\
Since this study is simply meant to evaluate the impact of the duty cycle emission from QSO, we did not rerun the Lyman-$\alpha$ simulation from \citet{21SSD}. We simply add the QSO duty cycle contribution to the local Ly-$\alpha$, and compute it with a simple approximation described below, similar to what 21cmFAST does \citep{Mesinger2011}.
\subsection{Lightcones} 
LICORICE outputs lightcones of the quantities relevant for $\delta T_b$ calculation (ionizing fraction, density, velocity and temperature). The lightcones are constructed on the fly, during the simulation. The observer is placed at the center of the box, looking in the $x,y$ or $z$ direction. We use the flat sky approximation, and assume that all lines of sight are parallel. Every time step, the relevant thin slice of the simulation box is added to the lightcone. Finally, we use periodic conditions such that the box is replicated about 6 times in the complete lightcone (from $z=15$ to $z=6$). For a more detailed discussion on the reasons and implications of these choices, see \cite{zawada2014}.\\
Finally, since $x_\alpha$ is obtained as post-treatment on the snapshots of the simulation, we cannot construct its lightcone the same way. We build it then by interpolating between thick slices whose thickness corresponds to the 10 Myr interval between the snapshots.
Interpolation is needed as one cannot assume $x_\alpha$ to be constant on a $10\,$Myr time scale, as shown in Figure 2 of \cite{zawada2014}.
\section{Quasar model}
\label{sectionmodelquasar}
In order to estimate the contribution of quasar duty cycle emission to the 21cm signal, we first need a model with which to assign a luminosity to a quasar for a given halo mass and redshift. In this section, we will present a simple model for computing the $\SI{1450}{\angstrom}$ luminosity, as well as the radio luminosity (at $\SI{1.4}{\giga\hertz}$). We will then compare our model with observed quasar luminosity functions (hereafter QLF). 
\subsection{Computing the optical intensity of a given quasar}
Let us consider a halo of mass $M_{\text{halo}}$ at a redshift $z$. According to the model of \cite{WyitheLoeb2003}, and for our choice of cosmology, its center contains a black hole of mass:\\
\begin{equation}
 M_{\text{BH}}=6.9\times 10^5 \left ( \frac{M_{\text{halo}}}{10^{12}M_\odot}\right)^{\frac 53} (1+z)^{\frac 5 2} M_\odot\label{eqMBH}
\end{equation}
There is some evidence that high redshift quasars accrete close to the Eddington limit: at $z=6$, \cite{Willott2010} find that $f_\text{edd}$ follows a lognormal distribution with peak $f_\text{edd}=1.07$ and dispersion 0.28\,dex. We choose for simplicity $f_\text{edd}=1$, and we can express the bolometric luminosity of the quasar as:\\
\begin{align}
 L_{\text{bol}}=f_{\text{corr}}L_{\text{Edd}}&=f_{\text{corr}}\times 3.2\times 10^4 \times\left ( \frac{M_{\text{BH}}}{M_\odot}\right ) L_\odot \\
 &= 2.2\times 10^{10}f_{\text{corr}}\left ( \frac{M_{\text{halo}}}{10^{12}M_\odot}\right)^{\frac 53} (1+z)^{\frac 5 2} L_\odot
\end{align}
where $f_\text{corr}$ is an ad-hoc coefficient used to correct for all the uncertainties on different coefficients, whose values are not well constrained in literature (Eddington's ratio, mass of the black hole, bolometric conversions...). We choose $f_\text{corr}= 1$ for our fiducial model.\\

Now we wish to determine the optical luminosity. Bolometric corrections in the optical range are usually given for $\SI{1450}{\angstrom}$, $\SI{3000}{\angstrom}$ and $\SI{5100}{\angstrom}$. The absolute magnitude $M_{1450}$ is the most commonly used quantity, and thus the most convenient to compare with other works. In addition, it has the tightest relation with $L_\text{bol}$ (see \cite{Nemmen2010}). According to \cite{Runnoe2012}, the bolometric luminosity usually relates to $L_{\SI{1450}{\angstrom}}$ as:\\
\begin{equation}
\log(L_{\text{bol}})=4.86 +0.91\log(\lambda L_\lambda(\SI{1450}{\angstrom}))\label{eqLbol1450}
\end{equation}
where $L_{\text{bol}}$ and $\lambda L_\lambda(\SI{1450}{\angstrom})$ are in $\SI{}{\erg\per\s}$. We want to determine the absolute magnitude $M_{1450}$:\\
\begin{equation}
 M_{1450}=-2.5\log\left ( \lambda L_\lambda(\SI{1450}{\angstrom})/(4\pi d^2)\right) \label{M1450L1450}
\end{equation}
where $d=10pc$ is expressed in meters and $\lambda L_\lambda(\SI{1450}{\angstrom})$ in $\SI{}{\erg\per\s}$. Using the bolometric correction above (equation \ref{eqLbol1450}), and the expression for the bolometric luminosity, we can compute $M_{1450}$:\\
\begin{multline}
 \label{eq1450}M_{1450}= -2.75\log(f_\text{corr})+4.58\log\left (\frac{M_\text{halo}}{10^{12}M_\odot}\right)\\
 +6.88\log(1+z)+17.2
\end{multline}
\subsection{Quasar duty cycle}
In order to include quasars in our simulation, we need to know the fraction of time a quasar spends in active mode. This is called the duty cycle of a quasar and can be defined as the fraction of quasars that are emitting at a given time. This parameter is poorly constrained, since it seems to depend strongly on the redshift and luminosity of the quasar. For instance, \cite{Shankar2013} compute a duty cycle around 0.01 (without obscuration) at $z=1.45$, but \cite{Shankar2010} find a strong increase in duty cycle with redshift, reaching almost $f_\text{duty}=1$ at $z=6$. However, \cite{DeGraf2017}, dispute this statement, arguing that the duty cycle is also a decreasing function of luminosity. At $z=4$ and $L_{bol}>10^{44}\SI{}{\erg\per\s}$, they find a duty cycle around $0.1$. However, the quasars that we are most interested in are in the $10^{44}-10^{48}\SI{}{\erg\per\s}$ range, so we might expect even lower duty cycles. \cite{Haiman2004}, in their quasar model, assumed that $f_\text{duty}=t_q/t_H(z)$, where $t_q=20$Myr and $t_H(z)$ is the hubble time at time $z$. This yields $f_\text{duty}=0.006$ at $z=2$ and $f_\text{duty}=0.021$ at $z=6$.\\

In our model we use $f_\text{duty}$ as a free parameter in order to reproduce observed quasar luminosity functions. We find that for $f_\text{corr}=1$ (no correction), $f_\text{duty}=0.02$ (independant of redshift) reproduces the observational luminosity functions well. On the contrary, $f_\text{duty}=t_q/t_H(z)$ reproduces the observations at $z=5-6$, but tends to underestimate the luminosities at $z=3.2$ and $z=4$. In order to avoid overestimation of the QLF at even higher $z$ ($6<z<12$) we opt for a constant duty cycle.\\

In addition, we will also show results with $f_\text{duty}=0.2$. Since we want to show the impact of changing the duty cycle while keeping the luminosity function constant, we will choose a corrective factor such that the two models predict the same luminosity function for the quasars in our simulation box. We find that using $f_\text{corr}=0.2$ together with $f_\text{duty}=0.2$, both model predict similar luminosity functions in our box, but this comes with a cost: the second model does not match the observations as closely.
\subsection{Optical luminosity function}
\label{subsecQLF}
We now wish to compute the optical quasar luminosity function for our model. That is the comoving number density of quasars per unit of magnitude, usually at $\lambda = \SI{1450}{\angstrom}$ (which correspond to the optical band on earth).\\

We use HMFs from cosmological simulations (\cite{Jenkins2001}) generated using HMFCalc (\cite{HMFCalc}). Along with the mass/luminosity relation computed above, this allows us to compute the luminosity function at any redshift.\\

\begin{figure}
 \centering
 \includegraphics[width=0.99\columnwidth]{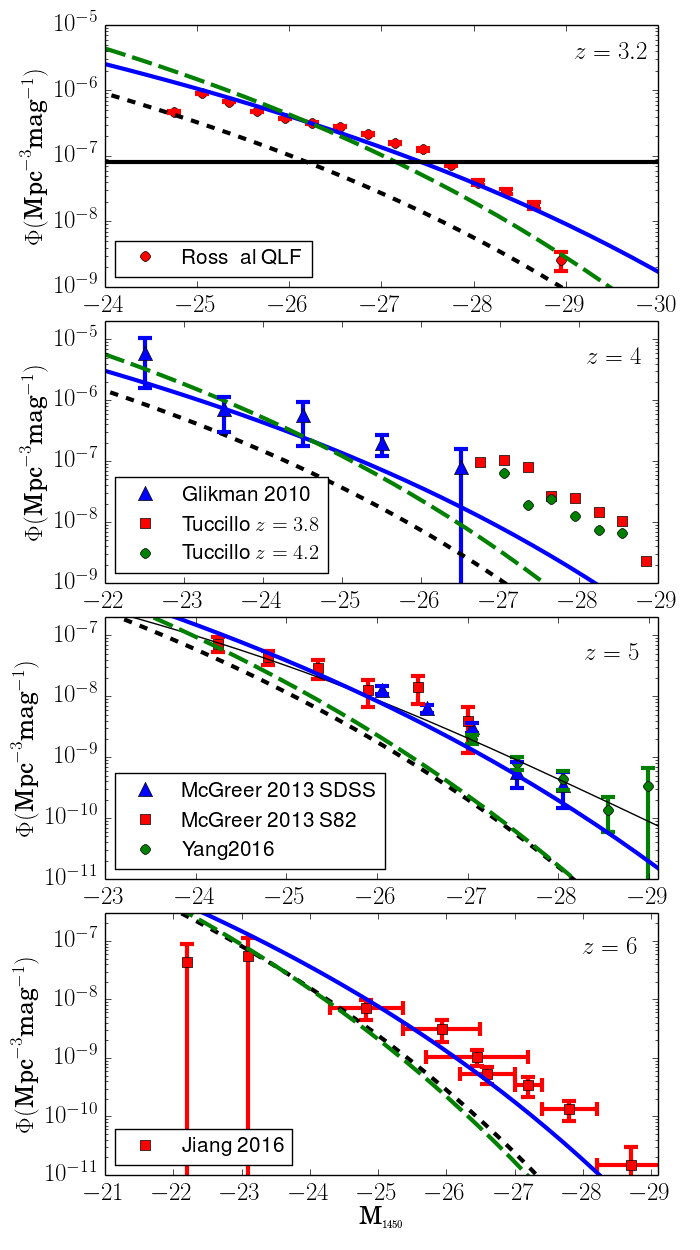}
 \caption{Quasar luminosity functions for the $\SI{1450}{\angstrom}$ absolute magnitude $M_{1450}$. The blue solid line represent our fiducial model with $f_\text{duty}=0.02$ and $f_\text{corr}=1$, the long dashes green line represent our model with $f_\text{duty}=0.2$ and $f_\text{corr}=0.2$ and the short dashed black line is the model of \protect\cite{Haiman2004}. Top panel: the theoretical curves are generated at $z=3.25$, and the data points are taken from \protect\cite{Ross2013} and correspond to the $3<z<3.5$ bin. Second from top: curves generated at $z=4$, blue squares taken from \protect\cite{Glikman2010} for $z\sim 4$, red and green squares taken from \protect\cite{Tuccillo2015} at respectively $z\sim3.8$ and $z\sim 4.2$. Third down from top: curves generated at $z=5$, blue and red squares  from \protect\cite{McGreer2013} at $z\sim 4.9$, respectively from SDSS main and Stripe 82, and green squares from \protect\cite{Yang2016} at $z\sim 5.05$. Bottom panel: curves generated at $z=6$, red squares taken from \protect\cite{Jiang2016} at $z\sim 6$.}
 \label{QLFFull}
\end{figure}

In figure \ref{QLFFull}, we produce luminosity functions at $z=3.25$, 4, 5 and 6 as predicted in our model, and compare them with observed QLF from literature. As we can see, we have reasonable agreement, except perhaps with \cite{Tuccillo2015}. As a general rule, we can notice that the slope of our models are slightly too high, and this increases with higher $f_\text{duty}$. We could reduce the slope by including a dispersion in equation \eqref{eq1450}. There are indeed many reasons to do so: almost all relations leading to our final equation have dispersion. According to \cite{Runnoe2012}, the bolometric correction in equation \eqref{eqLbol1450} has rather small scatter, but as we can see in \cite{Willott2010}, the Eddington ratio is expected to show larger scatter. In addition, the relation between the mass of the halo and that of the black hole (equation \ref{eqMBH}) is more of a scaling law, and we might expect significant scatter in the values of the constant and exponent in the equation. However, given how little we know about high redshift quasars, and how well our model reproduces observations, we choose not to increase our model's complexity.\\
We insist on the fact that, according to our model, the quasars in our simulation box or in a typical SKA field have lower luminosities than observations have reached. One can consider that the contributing quasars in our simulation are those above $\phi(M)=10^{-5}\text{Mpc}^{-3}\text{mag}^{-1}$ in figure \ref{QLFFull}, a value that corresponds, for our simulation volume, to approximately one quasar per unit redshift.\\

This means that the quasars we expect in our simulation volume are almost completely out of the luminosity range of the observed QLF. As a result, the three models presented in figure \ref{QLFFull} ($f_\text{duty}=0.02$, $f_\text{duty}=0.2$ and \cite{Haiman2004}), even though they do not reproduce the observations equally well, predict very similar quasar luminosity functions in our simulation box.

Using our model we can also comment on the recent discovery of a $z\sim 7.5$ quasar (\cite{BanadosQuasar}), with absolute magnitude $M_{1450}\simeq -27.86$. The authors searched approximately $\SI{2500}{\deg\squared}$ for quasars with $z>7$ and $M_{1450} \le -26.2$ (the UKIDSS flux limit). With those assumptions, our model predicts an average of 0.7 quasars, which is consistant with this discovery.

\subsection{Distribution of radio loudness}
In the previous subsection, we modelled the relation between the mass of the halo and the luminosity at $\SI{1450}{\angstrom}$. Unfortunately, there is no tight relation between the UV luminosity and radio luminosity. We define the radio loudness of a given quasar as the logarithm of the ratio of the spectral flux densities at $\SI{1.4}{\giga\hertz}$ and in the SDSS-$i$ band ($\lambda_\text{eff}=\SI{7471}{\angstrom}$). That is:
\begin{equation}
 R=\log(F_\nu(\SI{1.4}{\giga\hertz})/F_\nu(\nu_i))
\end{equation}
where $\nu_i=c/\SI{7471}{\angstrom}\simeq \SI{4e14}{\hertz}$, and $F_\nu$ is the spectral flux density of  the quasar. According to \cite{Balokovic2012}, at high redshift $(2<z<5)$, the distribution of radio loudness follows a double gaussian distribution. Around 10\% of the quasars belong to the ``radio loud'' population, whose radio loudness follows a normal distribution:
\begin{equation}
 P(R) = \frac 1{\sqrt{2\pi\sigma^2}} e^{-\frac 12\frac{(R-R_0)^2}{\sigma^2}},~R_0 \simeq 1.95,~ \sigma\simeq 0.87\label{RLdistrib}
\end{equation}
The trend is for this dispersion to tighten as $z$ increases, and the mean radio loudness seems to increase as well. However, as a simple model, we will assume this relation to hold for higher redshifts (from $z\simeq 6$ to $z\simeq 12$), where we know nothing from observations.\\

In order to go from the bolometric spectral density of luminosity to the $\SI{7471}{\angstrom}$ spectral density of luminosity, we chose the $\SI{5100}{\angstrom}$ bolometric conversion from \cite{Runnoe2012}, which is closer than the one we used to compute the QLF (equation \ref{eqLbol1450}). Then:
\begin{equation}
 \log (L_\text{bol})=4.76+0.91\log(\lambda F_\lambda(\SI{5100}{\angstrom}))
\end{equation}
We finally use the template from \cite{Atlas2011} in order to compute the conversion from $\SI{5100}\angstrom$ to $\SI{7471}{\angstrom}$ :
\begin{equation}
 \lambda F_\lambda(\SI{7471}{\angstrom}) \simeq 0.78\lambda F_\lambda (\SI{5100}\angstrom)
\end{equation}
We could also have used a fixed spectral index, and the result would have been similar.\\

Finally, we would like to add that according to, e.g., \citet{Sikora2007}, the quasars with a high radio loudness parameter tend to have a low Eddington ratio (and conversely), that means that our assumption of an Eddington ratio being close to 1 may not hold for radio loud quasars. However, as pointed out earlier, the Eddington ratio is degenerate with other quantities and we will judge our model on its capacity to predict the observed radio QLF.
\subsection{Radio luminosity functions}
\label{subsecrqlf}
\begin{figure}
 \centering
 \includegraphics[width=\columnwidth]{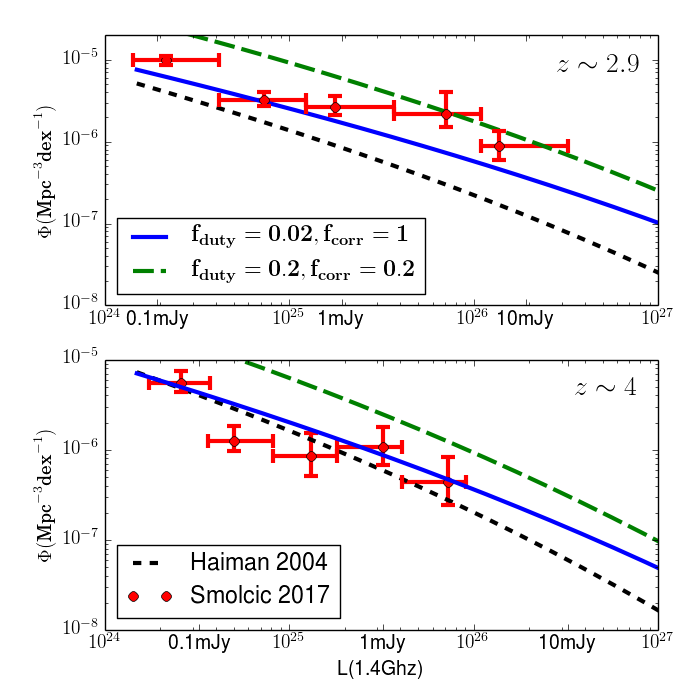}
 \caption{Radio quasar luminosity functions. The blue solid line represents our fiducial model with $f_\text{duty}=0.02$ and $f_\text{corr}=1$, the dashed green line represents our model with $f_\text{duty}=0.2$ and $f_\text{corr}=0.2$ and the short dashes black line is a comparison with \protect\cite{Haiman2004}. Red circles are RQLF from \protect\cite{Smolcic2017}. Top: $z\sim 2.9$, bottom: $z\sim 4$}
 \label{RQLFFull}
\end{figure}

We now only need to convolve the normal distribution given above with the luminosity functions that we have computed in subsection \ref{subsecQLF}, in order to generate the radio luminosity function. The results are shown in figure \ref{RQLFFull}. Observations at the bright end of the radio quasar luminosity function are rare. We compare our results with \cite{Smolcic2017}, who computed the radio luminosity functions of AGNs in different redshift bins. In particular, figure \ref{RQLFFull} shows the comparaison between our model and the two highest redshift bins, $3.5<z<5.5$, ($z_{\text{MED}}=4$) and $2.5<z<3.5$, ($z_{\text{MED}}=2.9$). We can see that our fiducial model $(f_\text{duty}=0.02)$ has a reasonable agreement for $z\sim 4$, but that we are a factor of 3 below the $z=2.9$ radio QLF.\\

However, according to \cite{Ghisellini2014}, above redshift $3-5$, the photons from the CMB partly suppress the synchrotron flux for extended radio sources. This is not included in our model and may explain that we are somewhat faint at low redshift but in agreement with observation at $z \sim 4$. Then it is possible that our model over-estimates the RQLF at $z > 6$ (when we do not have observations).

\subsection{Predictions relevant for the 21-cm forest}
Using our model, we now want to predict the abundances of radio loud quasars in a typical SKA field. For that, we will consider that the SKA field of view is $\SI{5}{\deg}\times \SI{5}{\deg}=\SI{25}{\deg\squared}$ wide. Also, the SKA low is located in Australia, at $28^\circ$ below the equator. If we assume that it can see up to $30^\circ$ over the horizon, then the available sky has a solid angle of $\SI{31500}{\deg\squared}$.\\

\begin{figure}
 \includegraphics[width=\columnwidth]{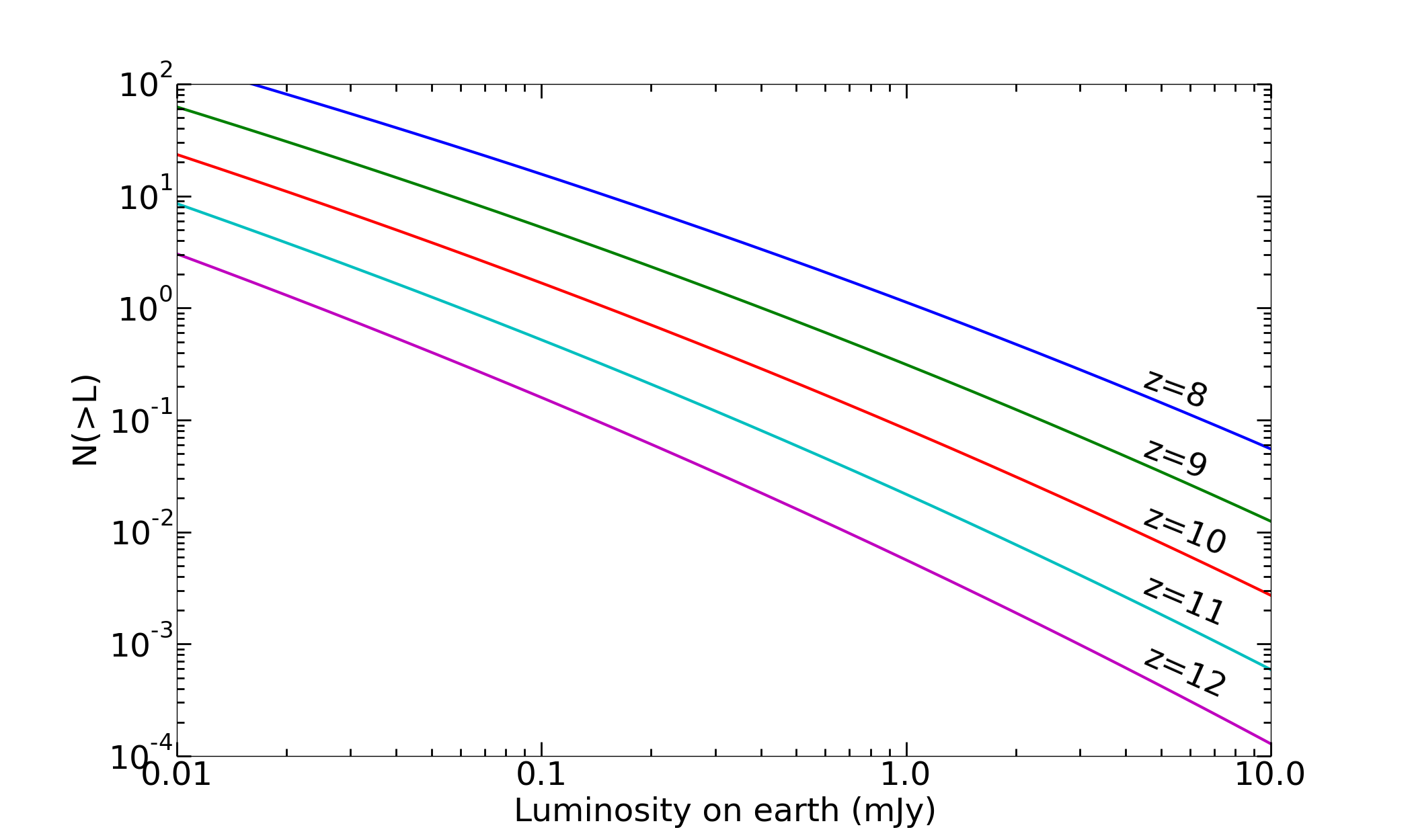}
 \caption{Average number of quasars above a given mJy luminosity that are expected in an SKA-like field of view, for different redshift bins. Redshift bins are centered at $z=8$, 9, 10, 11 and 12, and have a width $\Delta z=1$.}
 \label{mJyQLF}
\end{figure}

In figure \ref{mJyQLF} we plot the number of quasars expected in an SKA-like field of view, above a given flux density and for different redshift bins. Also, in table \ref{tableprobmJy} we record the number of $\SI{1}{\mJy}$ and $\SI{10}{\mJy}$ quasars that our model predicts for an SKA-like field of view, and for the fraction of the sky accessible to SKA-Low. As we can see, according to our model, there is little chance to find even a single $\SI{1}{\mJy}$ quasar at $z \ge 9.5$ in a typical SKA field. However, in the full SKA sky we can expect at $z\ge9.5$ several hundred quasars over $\SI{1}{\mJy}$, and even a few over $\SI{10}{\mJy}$. These results suggest that one shouldn't expect to find a suitable quasar for the 21cm forest in a single SKA field chosen, for example, to perform 21-cm tomography in optimal conditions. However, if the shallow survey described in \cite{Koopmans2015} (section 6.3) (over $\SI{10000}{\deg\squared}$, that is 1/3 of the SKA available sky) was conducted beforehand, a bright ($> 10$ mJy) radio loud quasar may be found and observed for 21-cm forest detection, at the same time as the tomography is performed on another field, using multi-beaming. This prediction, again, does not account for the possible suppression of the synchrotron flux by the CMB.
\begin{table}
\centering
\begin{tabular}{|c|c|c|c|}
\hline
Redshift & [12.5,11.5]& [11.5,10.5]& [10.5,9.5] \\
\hline
$N_\text{field}(F>\SI{1}{\mJy})$ & $5.8\times 10^{-3}$& $2.2\times 10^{-2}$& $8.4\times 10^{-2}$ \\
\hline
$N_\text{field}(F>\SI{10}{\mJy})$ & $1.3\times 10^{-4}$ & $6.0\times 10^{-4}$ & $2.7\times 10^{-3}$\\
\hline
$N_\text{SKY}(F>\SI{1}{\mJy})$ & 7.4 & 28 & 106 \\
\hline
$N_\text{SKY}(F>\SI{10}{\mJy})$ & 0.2 & 0.9 & 3.8\\
\hline
\end{tabular}
\caption{Average number of quasars above $\SI{1}{\mJy}$ or $\SI{10}{\mJy}$, for one SKA-like field of view (around $\SI{25}{\deg\squared}$) or in the available sky for the SKA (around $\SI{31500}{\deg\squared}$), for different redshift bins.}
\label{tableprobmJy}
\end{table}

\section{Influence of a radio-loud quasar on the 21cm signal from the neighbouring IGM}
\label{section21cm}
\subsection{Computing the spin temperature with a radio loud QSO contribution}
A detailed analysis of the 21-cm signal and $\delta T_b$ calculation can be found in \cite{ReviewFurlanetto}. In this subsection we will only rewrite the main equations needed to take the QSO contribution into account. Neutral hydrogen in the IGM is subject to several processes that regulate the population of the ground state hyperfine levels. Spontaneous emission ($A_{10}$), stimulated emission ($B_{10}$) and absorption $(B_{01})$ of 21cm background radiation (from both CMB and radio loud quasars) are effective processes. Also, one has to consider pumping by lyman-$\alpha$ photons ($P_{10}$ and $P_{01}$) as well as collisions ($C_{10}$ and $C_{01}$). Taking all these into account, the variation of the hyperfine levels population is:
\begin{multline}
 \label{eqdn0dt}\frac{dn_0}{dt} = n_1 \left ( A_{10} + \left ( 4\pi I_\text{CMB}+I_\text{QSO}\right ) B_{10} + P_{10} + C_{10} \right) \\
 - n_0 \left ( \left ( 4\pi I_\text{CMB}+I_\text{QSO}\right)B_{01} + P_{01} + C_{01}\right)
\end{multline}
where $n_0$ is the number of atoms in the ground level, $n_1$ is the number of atoms in the first hyperfine level, $I_\text{CMB}$ is the spectral intensity of the CMB per unit of solid angle, and $I_\text{QSO}$ is the spectral intensity of the nearby quasars. $I_\text{QSO}$ is usually omitted in this relation, as the CMB is assumed to be the sole source of background 21cm radiation. However, this is precisely what we want to explore: whether the radio luminosity of the quasar substantially changes the value of $\delta T_b$.\\

The expression above can be simplified using usual relations between Einstein's coefficients $A_{10}=\frac{8\pi h\nu^3}{c^2}B_{10}$, $B_{01}=3B_{10}$ and the Raleigh-Jean's relation $I_\text{CMB}=\frac{2\nu^2k_B}{c^2}T_\text{CMB}$. We will also make the assumption that the IGM is in equilibrium, and that $T_\text{21cm}\simeq 0.068K$ is small compared to $T_K$, $T_C$ and $T_S$. We can then write $\frac{n_1}{n_0} \simeq 3\left (1-\frac{T_{\text{21cm}}}{T_S}\right)$, $C_{01}=3\left ( 1-\frac{T_{\text{21cm}}}{T_K}\right)C_{10}$ and $P_{01}=3\left ( 1 - \frac{T_{\text{21cm}}}{T_C}\right)P_{10}$. One can then express equation \ref{eqdn0dt} as:
\begin{equation}
 T_S^{-1} = \frac{T_{\text{rad}}^{-1} + x'_\alpha T_C^{-1} + x'_C T_K^{-1}}{1+x'_\alpha+x'_C}
\end{equation}
with
\begin{equation}
 T_{\text{rad}} = T_{\text{CMB}} + \frac{c^2}{8\pi\nu^2 k_B} I_\text{QSO}
\end{equation}
and
\begin{equation}
 x'_\alpha =\frac{P_{10}T_{\text{21cm}}}{A_{10}T_{\text{rad}}} ~~\text{and}~~x'_C=\frac{C_{10} T_{\text{21cm}}}{A_{10}T_{\text{rad}}}
\end{equation}
One can finally compute the differential brightness temperature of the 21-cm signal, taking into account the effect of QSOs.
\begin{multline}
 \delta T_b=27 x_n (1+\delta) \left ( \frac{1+z}{10}\right)^{\frac 12} \frac{T_S-T_{\text{CMB}}}{T_S}\left (1+\frac 1{H(z)} \frac{dv_{||}}{dr_{||}}\right)^{-1} \\
 \times\left (\frac{\Omega_b}{0.044}\frac h{0.7}\right)\left(\frac{0.27}{\Omega_m}\right)^{\frac12} \left ( \frac{1-Y_p}{1-0.248}\right) \SI{}{\milli\kelvin}\label{dtb}
\end{multline}
Computationally speaking, the changes are very small and mainly consists of replacing $T_{\text{CMB}}$ with $T_{\text{rad}}$ everywhere except in equation \ref{dtb} (indeed, the signal is still observed in absorption or emission against the CMB). Also, if $x_\alpha$ is already computed, it has to be replaced with $x'_\alpha = x_\alpha \times T_{\text{CMB}}/T_{\text{rad}}$. We can compute the local flux received from the quasar with a simple $r^{-2}$ law assuming negligible absorption and considering that the effect is non-negligible only at distances much smaller than the Hubble radius.
\subsection{Influence of a $\mathbf{> 1}$ mJy quasar on the tomography}
\label{sectionbigquasars}
As we saw in section \ref{subsecrqlf}, the observation of a $\SI{10}{\mJy}$, or even a $\SI{1}{\mJy}$ quasar between $z=10$ and $12$ would require a survey over a substantial area of the sky (see table \ref{tableprobmJy}). However, such bright quasars are usually considered for example as background sources for $21$-cm forest studies \citep[e.g.][]{Semelin2016, Ciardi2013}, and for that reason, in this section we consider the effect of such quasars on the 21-cm tomography. More precisely, using a FoF halo finder on the dark matter, we choose three of the biggest haloes in our simulation. We place a quasar in their center and assign it a flux density of $\SI{1}{\mJy}$ and $\SI{10}{\mJy}$. Our haloes have redshift 11.7, 10.7 and 10.2 and masses of $1.1\times 10^{11}\text{M}_\odot$, $1.3\times 10^{11}\text{M}_\odot$ and $3.2\times 10^{11}\text{M}_\odot$.\\

As we have shown, $\delta T_b$ is sensitive to the instantaneous value of $I_\text{QSO}$. For that reason, the result is be affected by the timing of the duty cycle. Indeed, the position of the quasar in the lightcone corresponds to a specific time $t_\text{QSO}$. Of course, if the quasar starts emitting at $t_\text{start}> t_\text{QSO}$, then the influence of the quasar will not be imprinted on the lightcone, and thus not be witnessed on earth. But even if we are able to witness the quasar beginning its duty cycle, if the quasar stops emitting at $t_\text{end}> t_\text{QSO}$, the end of the emission will not leave an imprint on the lightcone. We will study two scenarios, the first one, called ``timing 1'', corresponds to a quasar starting its duty cycle $\Delta t_\text{start}=\SI{1.8}{\mega\yr}$ before the time at which it is seen in the lightcone, and emitting for $t_\text{duty}=\SI{2}{\mega\yr}$. The second one, ``timing 2'', corresponds to $\Delta t_\text{start}=\SI{2.4}{\mega\yr}$ and $t_\text{duty}=\SI{2}{\mega\yr}$.

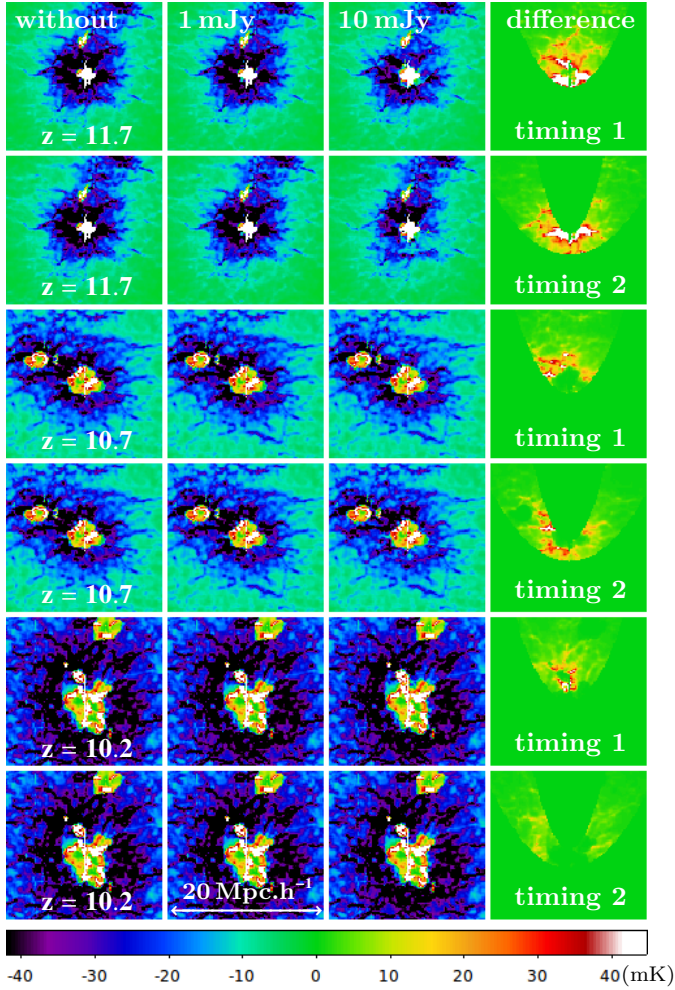
\begin{figure}
 \begin{picture}(220,370)
\put(0,0){\includegraphics[width=\columnwidth]{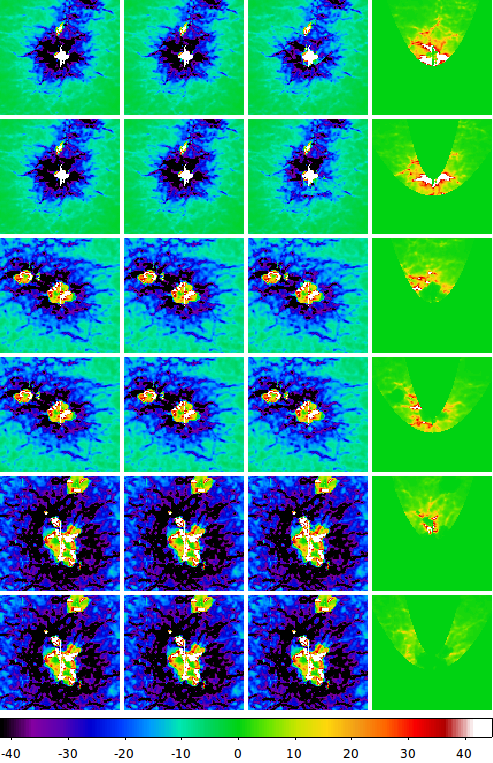}}
\put(13,29){\large $\color{white}\bf z = 10.2$}
\put(190,31){\textcolor{white}{\large \bf timing 2}}
\put(13,87){\large $\color{white}\bf z = 10.2$}
\put(190,89){\textcolor{white}{\large \bf timing 1}}
\put(13,145){\large $\color{white}\bf z = 10.7$}
\put(190,147){\textcolor{white}{\large \bf timing 2}}
\put(13,203){\large $\color{white}\bf z = 10.7$}
\put(190,204){\textcolor{white}{\large \bf timing 1}}
\put(13,261){\large $\color{white}\bf z = 11.7$}
\put(190,262){\textcolor{white}{\large \bf timing 2}}
\put(13,318){\large $\color{white}\bf z = 11.7$}
\put(190,320){\textcolor{white}{\large \bf timing 1}}
\put(3,362){\textcolor{white}{\large \bf without}}
\put(64,362){\textcolor{white}{\large \bf 1\,mJy}}
\put(124,362){\textcolor{white}{\large \bf 10\,mJy}}
\put(187,362){\textcolor{white}{\large \bf difference}}
\put(230,2.5){(mK)}
\put(61,27){\boldmath$\color{white}\xleftrightarrow{\mathmakebox[1.85cm]{\text{\normalsize\bf 20\,Mpc.h}^{-1}}}$}
\end{picture} 
 \caption{Effect on the 21-cm signal of the radio emission from three of the biggest quasars expected in a standard SKA field, at different redshifts and for different duty cycle timings. From left to the right: $\delta T_b$ (in mK) without a quasar, with a $\SI{1}{\mJy}$ quasar, with a $\SI{10}{\mJy}$ quasar, and the difference between the third image and the first (that is, the net contribution of the $\SI{10}{\mJy}$ quasar).}
 \label{bigquasars21cm}
\end{figure}

Results can be found in figure \ref{bigquasars21cm}. As we can see, there are very specific patterns for the two different timings. Those patterns correspond to the intersection in space-time of the past lightcone of the $z=0$ observer with the future lightcones of the collection of events defined by the position of the quasar and all the instants during its duty cycle.  Depending on the timing, this pattern can be approximated (assuming a distance to the quasar much smaller than the distance to the observer) to one full paraboloid (timing 1), or a shell between two paraboloids (timing 2). This pattern has already been shown in  figure 7 of \cite{zawada2014}  for the lightcone of $x_\alpha$. One should also note that in a configuration such as ``timing 2'', the quasar could have an impact on the $\delta T_b$, but the source itself would not be observable, since it is was not active when its worldline intersected the past lightcone of the observer.\\

One can see that the effect can reach over $\SI{40}{\milli\kelvin}$ in the immediate vicinity of the quasar, and over $\SI{10}{\milli\kelvin}$ in extended regions. One can even see a distinct pattern at $z=11.7$. However, the observability of this phenomenon seems to decrease strongly with decreasing redshift. There are two main limiting factors: the first one is the intrinsically high $\delta T_b$ around the halo, which can mask the effect of the quasar radio emission. The other is the extension of the ionized bubbles inside which $\delta T_b$ is close to zero, regardless of the influence of the quasar. These two limiting factors are all the more daunting in that they mask the effect of the quasar where it is most prominent (in the immediate vicinity of the quasar). For redshifts below 10, those two effects almost completely mask the effect of the quasar, which is why we choose not to show quasars at $z=9$ and below.\\

Even though the effect looks quite important, there is little chance that it would be observable with the SKA. Indeed, we will see in subsection \ref{subsecresolxa} that, taking thermal noise into account, the second effect we present in this work barely reaches observability, even though its magnitude is considerably higher.
\subsection{All quasars}
Quasars whose radio contribution produced a difference greater than $\SI{10}{\milli\kelvin}$ in $21$-cm emission from the nearby IGM is exceedingly rare: there is less that a $1\%$ chance of finding one in an SKA-like field of view according to our model. In this section we assess the combined effect of all quasars actually present in the field. We first run the FoF halo finder on the different snapshots of the simulation, and then recreate a lightcone of haloes. Each halo in this lightcone then has a 10\% chance of hosting a radio loud quasar, and to these, we assign a $\SI{1450}{\angstrom}$ luminosity following equation \eqref{eq1450}. Then, we choose the radio loudness of each quasar stochastically following equation \eqref{RLdistrib}. Finally, we randomly choose a starting time for the duty cycle of our quasar, in the 100\,Myrs before the time when the worldline of the quasar intersects the past lightcone of the observer. For our fiducial model, with $f_\text{duty}=0.02$, the quasar will shine for $\SI{2}{\mega\yr}$, and for our model with $f_\text{duty}=0.2$, it will shine for $\SI{20}{\mega\yr}$. These are somehow arbitrary choices, and it does change the form of the contribution, but we expect the mean effect is not affected by varying the duty time, with constant $f_\text{duty}$.\\

\begin{figure*}
\begin{picture}(600,320)
 \centering
 \put(0,0){\includegraphics[width=\textwidth]{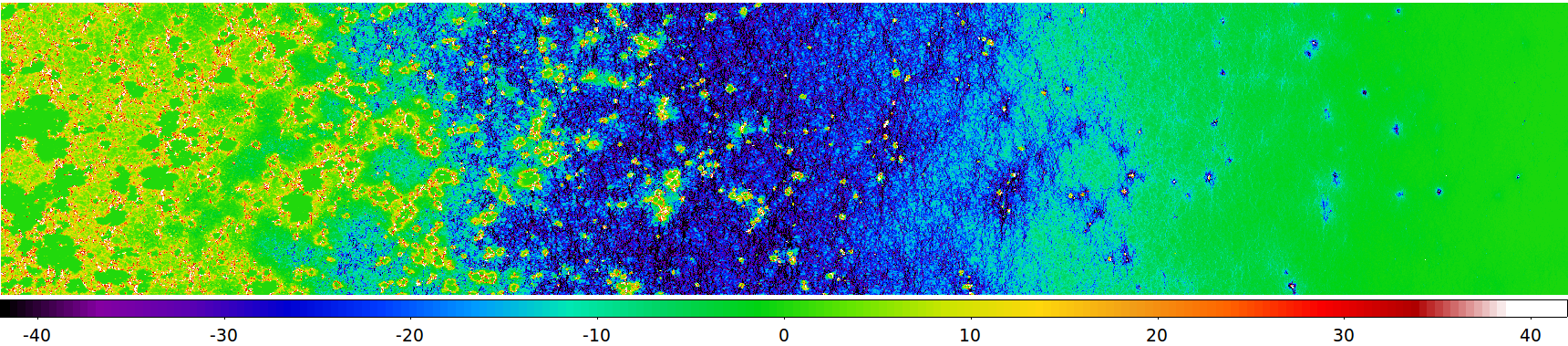}}
 \put(0,100){\includegraphics[width=\textwidth]{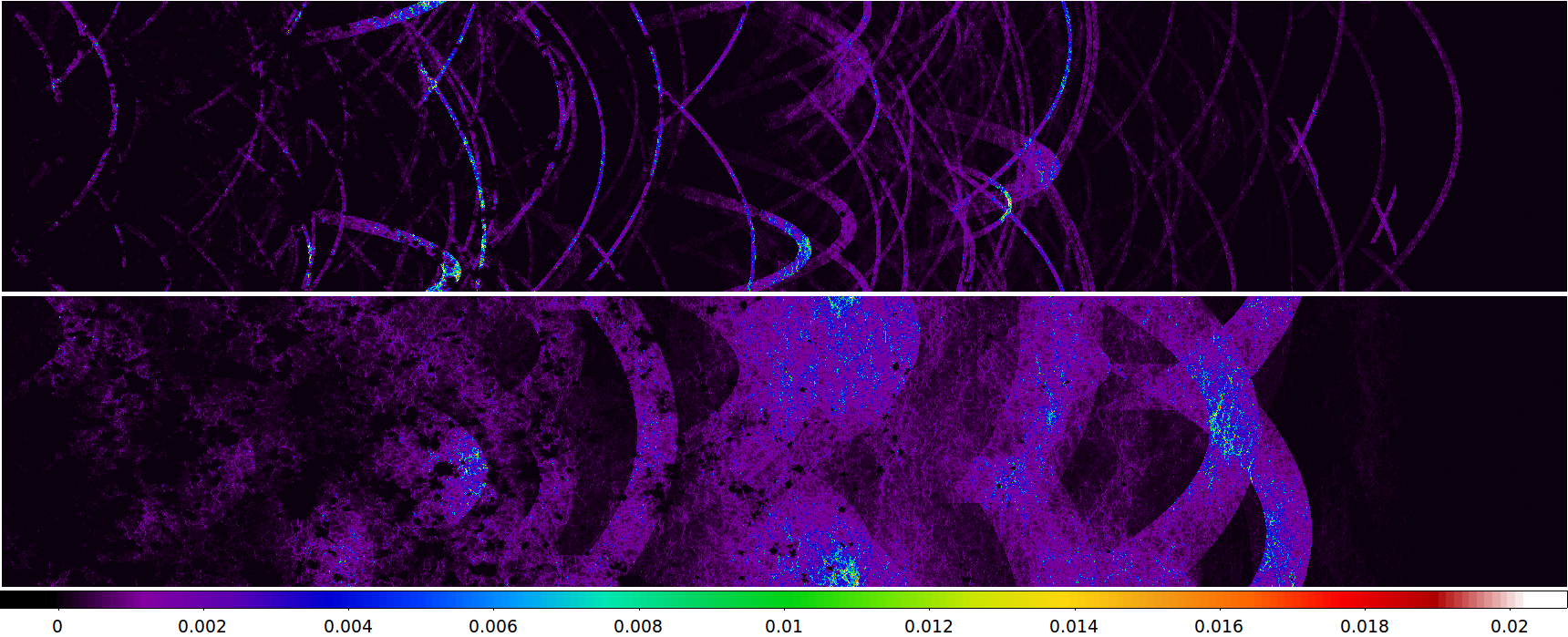}}
\put(498,56){$\xupdownarrow{1.55cm}$}
 \put(488,38){\rotatebox{90}{$200\text{Mpc}.h^{-1}$}}
 \put(414,122){{$\color{white}f_\text{duty}=0.2,~f_\text{corr}=0.2$}}
 \put(417,218){{$\color{white}f_\text{duty}=0.02,~f_\text{corr}=1$}}
 \put(495,1.3){(mK)}
 \put(491,101.3){(mK)}
 \put(0,315){$z$}
 \put(41.5,315){$7$}
 \put(42,306){\bf|}
\put(171.7,315){$8$}
 \put(172.2,306){\bf|}
\put(276.2,315){$9$}
 \put(276.7,306){\bf|}
\put(358.5,315){$10$}
 \put(360.9,306){\bf|}
\put(429.5,315){$11$}
 \put(432,306){\bf|}
\put(489.4,315){$12$}
 \put(492,306){\bf|}
 \end{picture}
 \caption{From top to bottom: Lightcone of $\delta T_b$ (in mK) without quasars, contribution of the quasars to $\delta T_b$ with $f_\text{duty}=0.02$ and $f_\text{corr}=1$, and contribution of the quasars to $\delta T_b$ with $f_\text{duty}=0.2$ and $f_\text{corr}=0.2$. The vertical axis is plotted with pixels of constant comoving distance, while the horizontal axis is plotted with pixels of constant expansion factor. As a result, pixels at $z=7$ have horizontal size 10\% bigger than vertical size, and 44\% bigger at $z=12$. }
 \label{allquasars21cm}
\end{figure*}
In figure \ref{allquasars21cm}, we can see the combined effect of all the quasars in the simulation. As we can see, the effect barely reaches 0.02\,mK in some places, and the effect on the mean is 10 times lower. This is far below the level of noise expected in the different versions of the SKA, and even if we showed a single slice of the lightcone, and if a particularly bright radio loud quasar could leave a stronger imprint, we expect a most likely undetectable contribution from this effect.
\section{Effect of the lyman continuum emission of the quasar}
\label{sectionxa}
\subsection{Computing the contribution of the quasar}
In this section, we will compute a simple model for the contribution of a quasar to the $x_\alpha$ coefficient:
\begin{equation}
 x_\alpha = \frac{4P_\alpha T_\star}{27A_{10}T_{\text{CMB}}}
\end{equation}
where $A_{10}=\SI{2.85e-15}{\per\s}$, $T_\star=\SI{0.068}{\kelvin}$ and $P_\alpha$ is the number of scatterings per atom per second. According to \citet{Semelin07}, the number of scatterings in the Lyman-$\alpha$ line per photon in the IGM at average density at redshift $z$ can be written:
\begin{equation}
N_{\text{scat}}=8\times 10^5 \frac{H(z=10)}{H(z)}
\end{equation}
Thus, for a given atomic density $n$, the number of scatterings per photon is:
\begin{equation}
 N_{\text{scat}}(n)=\num{8e5}\frac{H(z=10)}{H(z)} \frac{n}{n_{\text{mean}}}
\end{equation}
where $n_{\text{mean}}$ is the mean atomic density of the universe.\\

Let us now consider a quasar with a constant spectral density of luminosity $L_{\alpha\beta}$, between Ly-$\alpha$ and Ly-$\beta$. We will consider a simple model, where at a certain comoving distance $r$ from the quasar, the flux is:
\begin{equation}
 \frac{L_{\alpha\beta}}{4\pi r^2} \SI{}{\erg\per\s\per\hertz\per\meter\squared}
\end{equation}
According to \cite{Semelin2007} and \cite{Chuzhoy2007}, this is true only at distances above $\SI{10}{\mega\pc}$. Figure 5 in \cite{Semelin2007} shows that a simple $r^{-2}$ law tends to underestimate $P_\alpha$ by up to a factor of 3 near the quasar. In order to archieve better precision, the full lyman-$\alpha$ simulation should be run, with the quasar contribution taken into account. However, we trust that the overall properties of this effect will still be correct.\\

Let us consider a spherical volume element $dr dS$. We will now compute the number of scatterings per atom per second in the cell. The expansion factor and comoving distance are related through the formula:
\begin{equation}
 dr=\frac{c da}{H(a) a^2}
\end{equation}
If a photon is emitted at an expansion factor $a_0$, with a frequency $\nu$, then it will be received at a distance $r$, and corresponding expansion factor $a(r)$, with a frequency $\nu a_0/a(r)$. The photons that redshift into the Ly-$\alpha$ line in that cell were emitted with frequencies approximately between $\nu_\alpha a(r)/a_0$ and $\nu_\alpha a(r+dr)/a_0$, meaning a bandwidth of $\Delta \nu=\nu_\alpha a(r)^2 H(a(r))dr/(a_0c)$. In a time interval of emission $dt_\text{em}$, the number of photons redshifted into the Ly-$\alpha$ line in this volume element can be expressed as:
\begin{equation}
 N=\left (\frac{\nu_\alpha a(r)^2 H(a(r))dr}{a_0c}\right)\left ( \frac{L_{\alpha\beta}}{4\pi r^2 E_\alpha(r)}dS\right)dt_\text{em}
\end{equation}
where $E_\alpha(r)=h\nu_\alpha a(r)/a$ is the energy of the emitted photon. If we call $F_\alpha$ the number of photons that are redshifted into the Ly-$\alpha$ line in the volume element per unit time, and $dt_\text{rec}$ the reception time interval, we can write:
\begin{equation}
 F_\alpha dt_\text{rec}=\frac{\nu_\alpha a(r)^2 H(a(r))dV}{a_0c}\frac{L_{\alpha\beta}}{4\pi r^2 h\nu_\alpha}dt_\text{em} 
\end{equation}
Using the relation $dt_\text{rec}/a(r)=dt_\text{em}/a_0$, one can deduce the following expression:
\begin{equation}
 F_\alpha=\frac{\nu_\alpha a_0 H(a(r))dV}{c}\frac{L_{\alpha\beta}}{4\pi r^2 h\nu_\alpha}
\end{equation}
This gives us the total number of scatterings per atom per second in our volume element of:
\begin{align}
P_\alpha&= \frac{F_\alpha N_\text{scat}(n)}{ndV}\\
&= \SI{8e5}H(z=10)n_\text{mean}^{-1}\frac{a_0 L_{\alpha\beta}}{4\pi r^2 h}
\end{align}

Which means that, at distance $r$ from the quasar, we can express the contribution from the quasar to $x_\alpha$ as:
\begin{align}
 x_{\alpha,\text{qso}}&=\frac{8\times 10^5 a_{\text{qso}}H(z=10)L_{\alpha\beta} T_\star}{27\pi r^2h A_{10}T_\text{CMB}n_{\text{mean}}c}
\end{align}
In order to apply our model, we choose, as noted earlier, an $x_\alpha$ efficiency $f_\alpha =0.5$. Though it seems a sensible choice, it also increases the effect we are studying by lowering the background $x_\alpha$. However, our choice of a $r^{-2}$ profile underestimates the luminosity by a factor of $1.5-2$ between 1 and $\SI{10}{\mega\pc}$ of the quasar (see figure 5 in \cite{Semelin2007}). Also, according to \citet{Vonlanthen2011} and \citet{Pritchard2006}, taking into account only the scattering in the lyman-$\alpha$ line and not in higher order lines would result in underestimating the luminosity by a factor of 3. For those reasons, we expect that we do not to overestimate the relative contribution of the quasar in $x_\alpha$.
\subsection{Single quasar}
\label{subsecsinglequasarsxa}
We want to assess the effect of the lyman band emission of a single bright quasar on the tomography. In order to choose our quasar, we will determine the brightest quasars expected in an SKA-like field of view, or in the SKA available sky. More precisely, we will choose three quasars in the redshift bins 12.5-11.5, 11.5-10.5 and 10.5-11.5, such that there is on average exactly one quasar at least as luminous, in an SKA-like field of view ($\SI{25}{\deg\squared}$) or the SKA available sky ($\SI{31500}{\deg\squared}$).\\

According to our simple model, there is a one to one relation between the mass and the UV luminosity for each quasar. For a given mass $m$ and a given redshift interval $[z-\Delta z/2,z+\Delta z/2]$ the average number of active quasars of mass greater than $m$ on the past lightcone of the observer, is given by:
\begin{equation}
 n_\text{hmf}(M>m)V(z,\Delta z)f_{\text{duty}}\label{eqprob}
\end{equation}
where $n_\text{hmf}$ is the cumulative halo mass function and $V(z,\Delta z)$ is the comoving volume element given by:
\begin{equation}
 V(z,\Delta z) = \int_{z-\Delta z/2}^{z+\Delta z/2} cD_M(z)^2/H(z)dz \Omega
\end{equation}
where $D_M(z)$ is the comoving distance between redshift $z$ and redshift 0 and $H(z) = H_0\sqrt{\Omega_m(1+z)^3+\Omega_\Lambda}$. In order to determine the mass of our quasar, we only have to solve for $m$ in:
\begin{equation}
 n(M>m)V(z,\Delta z)f_{\text{duty}}=1\label{eqmprob}
\end{equation}
In table \ref{tablequasarsxa} we present the properties of these standard quasars, and the quasars we choose in the simulation to represent them. Even though we have around 10 times less volume in our simulation than in a typical SKA field, we can still find quasars with masses close to the standard masses calculated in 
\ref{eqmprob}. We simply force them to be in the active phase, thereby offsetting the much smaller volume of our box.

\begin{table}

\begin{tabular}{|>{\centering}m{0.33cm}|>{\raggedleft}m{0.3cm}|c|>{\centering}m{1.1cm}|>{\centering}m{1.1cm}|>{\centering}m{1.3cm}|c|}
\hline
$\Omega$ & \hspace*{-0.35em}$z_{bin}$ & z & $\text{M}_{\text{stand}}$ (${M_\odot}$)& $\text{M}_{\text{real}}$ (${M_\odot}$)& lum. (erg/s/Hz)& $\text{M}_{1450}$\\
\hline
\hspace*{-0.35em}field & 12 & 11.7 & $8\times10^{10}$ & $7\times 10^{10}$ & $1.4\times 10^{29}$&-19.9\\
\hspace*{-0.35em}field & 11 & 10.7 & $1.3\times10^{11}$ & $1.3\times 10^{11}$ & $2.9\times 10^{29}$&-20.6\\
\hspace*{-0.35em}field & 10 & 10.0 & $2.3\times10^{11}$ & $2.2\times 10^{11}$ & $7.1\times 10^{29}$&-21.5\\
\hspace*{-0.18em}sky & 12 & 11.7 & $4.7\times10^{11}$ & $7\times 10^{10}$& $4.9\times 10^{30}$&-23.4\\
\hspace*{-0.18em}sky & 11 & 10.7 & $7.4\times10^{11}$ & $1.3\times 10^{11}$ & $9.9\times 10^{30}$&-24.1\\
\hspace*{-0.18em}sky & 10 & 10.0& $1.3\times10^{12}$ & $2.2\times 10^{11}$& $2.45\times 10^{31}$&-24.9\\
\hline
\end{tabular}
\caption{Properties of the quasars used in subsection \ref{subsecsinglequasarsxa}. Their mass ($\text{M}_\text{stand}$), and thus their luminosity, are chosen such that there is, on average, $p$ quasars above this mass in a typical SKA1 field, for given redshift bin. Then we try to find a halo with a mass ($\text{M}_\text{real}$) as close as possible to $\text{M}_\text{stand}$ in the corresponding redshift bin (actual redshifts are given in the table). The redshift bins are centered on $z_\text{bin}$ and have a width $\Delta z=1$.}
\label{tablequasarsxa}
\end{table}

\begin{figure}
\begin{picture}(210,495)
\put(0,0){\includegraphics[width=\columnwidth]{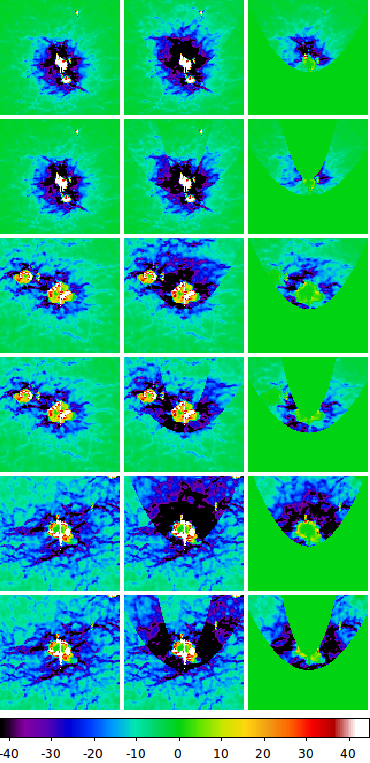}}
\put(23,40){\large $\color{white}\bf z = 10.2$}
\put(181,40){\large \color{white}\bf timing 2}
\put(23,117){\large $\color{white}\bf z = 10.2$}
\put(181,117){\large \color{white}\bf timing 1}
\put(23,193){\large $\color{white}\bf z = 10.7$}
\put(181,193){\large \color{white}\bf timing 2}
\put(23,270){\large $\color{white}\bf z = 10.7$}
\put(181,270){\large \color{white}\bf timing 1}
\put(23,348){\large $\color{white}\bf z = 11.7$}
\put(181,348){\large \color{white}\bf timing 2}
\put(23,426){\large $\color{white}\bf z = 11.7$}
\put(181,426){\large \color{white}\bf timing 1}
\put(18,485){\large \color{white}\bf without}
\put(106.5,485){\large \color{white}\bf with}
\put(175,485){\large \color{white}\bf difference}
\put(82,37){\boldmath$\color{white}\xleftrightarrow{\mathmakebox[2.45cm]{\text{\normalsize\bf 20Mpc.h}^{-1}}}$}
\put(232,4){(mK)}
\end{picture} 
 \caption{Effect of the lyman continuum emission of three of the biggest quasars expected in a typical SKA field on the 21-cm signal, at different redshifts and for different timing of the duty cycle. From left to right: $\delta T_b$ (in mK) without a quasar, with a standard quasar and the difference between the third image and the first (that is, the net contribution of the quasar).}
 \label{bigquasarsxa}
\end{figure}
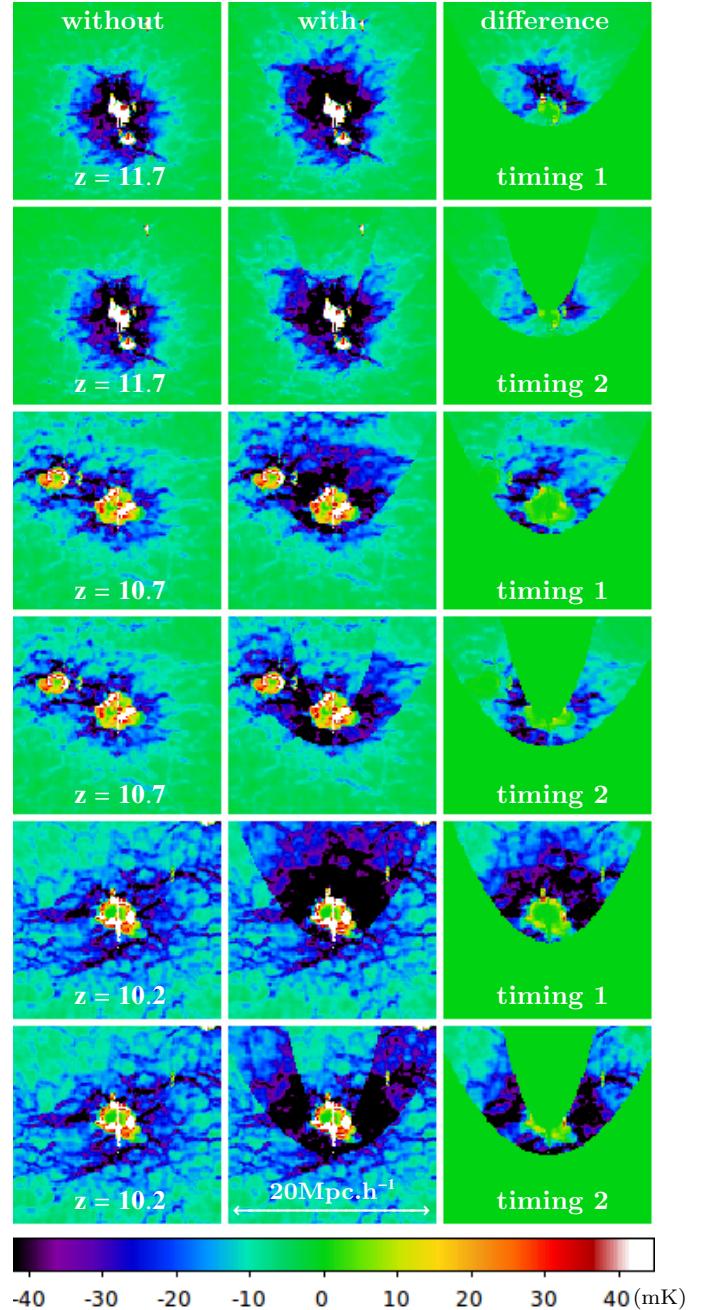

In figure \ref{bigquasarsxa}, we can see the results of this study. As can be seen, we have a strong impact on the tomography, over $\SI{-40}{\milli\kelvin}$ in extended regions. Also the patterns are clearly distinct from anything that is otherwise expected in the tomography. The effect is reduced at higher redshift, probably due to lower intensity, and it is at its peak around $z=10.2$. For lower redshift ($z=9$ and less) the effect almost completely disappears due to a higher mean $x_a$.
\subsection{Observability at SKA1 resolution}
\label{subsecresolxa}
In subsection \ref{subsecsinglequasarsxa}, we showed that for a typical SKA field of view, we should expect to see at least one quasar with significant contribution to $\delta T_b$ (extended regions of $\Delta \delta T_b \ge \SI{40}{\milli\kelvin}$). However, at SKA-like resolution, it is not obvious that the effect can be spotted in the tomography, since the shape of the pattern will be significantly altered by the resolution and noise. In order to address this question, we choose the $z=10$ quasar in table \ref{tablequasarsxa}, and we assign it a luminosity corresponding to the brightest quasar expected in an SKA-like field of view, and in the SKA available sky. 

Then, we considered a 1000h observation (in runs of 8h centered on the zenith) with the SKA, using the baseline design station layout\footnote{\href{https://astronomers.skatelescope.org/wp-content/uploads/2016/09/SKA-TEL-SKO-0000422\_02\_SKA1\_LowConfigurationCoordinates-1.pdf}{https://astronomers.skatelescope.org/wp-content/uploads/2016/09/SKA-TEL-SKO-0000422\_02\_SKA1\_LowConfigurationCoordinates-1.pdf}}. We assumed stations with a $35$\,m diameter, dipoles with effective collecting area $\mathrm{min}(2.56,{\lambda^2 \over 3})\,\text{m}^2$ and a system temperature $T_{\mathrm{sys}}=100+300\left({\nu \over 150 \mathrm{MHz}} \right)^{-2.55}\,\text{K}$ \citep{Mellema2013}. We evaluated the noise in each visibility cell following \citet{McQuinn2006}, computed the effective observation time in each cell resulting from earth rotation, and implemented the resolution choice with a gaussian filter in visibility space.
We generate the maps at three levels of resolution: the full resolution of our lightcone (without noise), and 6.8' and 3.4' where we added the thermal noise. We did that for two different timings (described in \ref{sectionbigquasars}).
\begin{figure}
 \begin{picture}(210,460)
 \put(62,458){\large Without quasar contribution}
 \put(30,446){\large 6.8'}
 \put(111,446){\large 3.4'}
 \put(181,446){\large unaltered}
 \put(28,352){\large Brightest quasar in an SKA-like field of view}
 \put(30,184){\large Brightest quasar in the SKA available sky }
 \put(0,364){\includegraphics[width=\columnwidth]{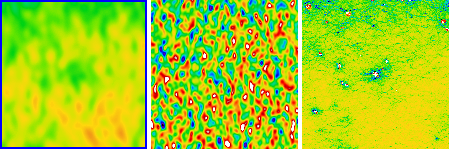}}
 \put(0,195){\includegraphics[width=\columnwidth]{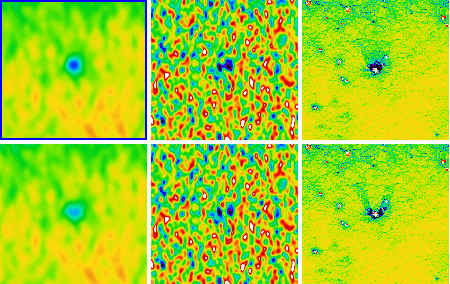}}
 \put(0,0){\includegraphics[width=\columnwidth]{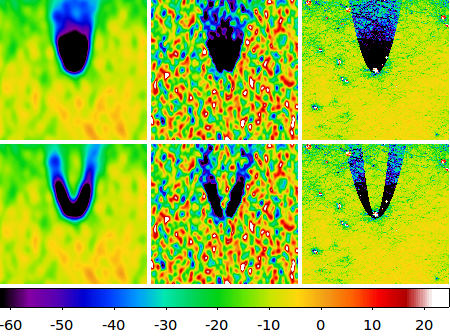}}
 \put(162,28){\boldmath$\xleftrightarrow{\mathmakebox[2.55cm]{\text{\normalsize\bf 120Mpc.h}^{-1}}}$}
 \put(232,4){(mK)}
 \end{picture}
 \caption{$\delta T_b$ maps (in mK) of one of the biggest quasars expected respectively in an SKA-like field of view or in the full SKA sky, for different timings and resolution. From top to bottom: image without the contribution of the quasar, image with the brightest quasar in an SKA-like field of view and timing 1, image with the brightest quasar in an SKA-like field of view and timing 2, image with the brightest quasar in the SKA sky and timing 1 and image with the brightest quasar in the SKA available sky and timing 2. From right to left, image without thermal noise, at the resolution of the simulation, image expected from the SKA at 3.4arcmin, and at 6.8arcmin.}
 \label{quasarsnoisy}
\end{figure}
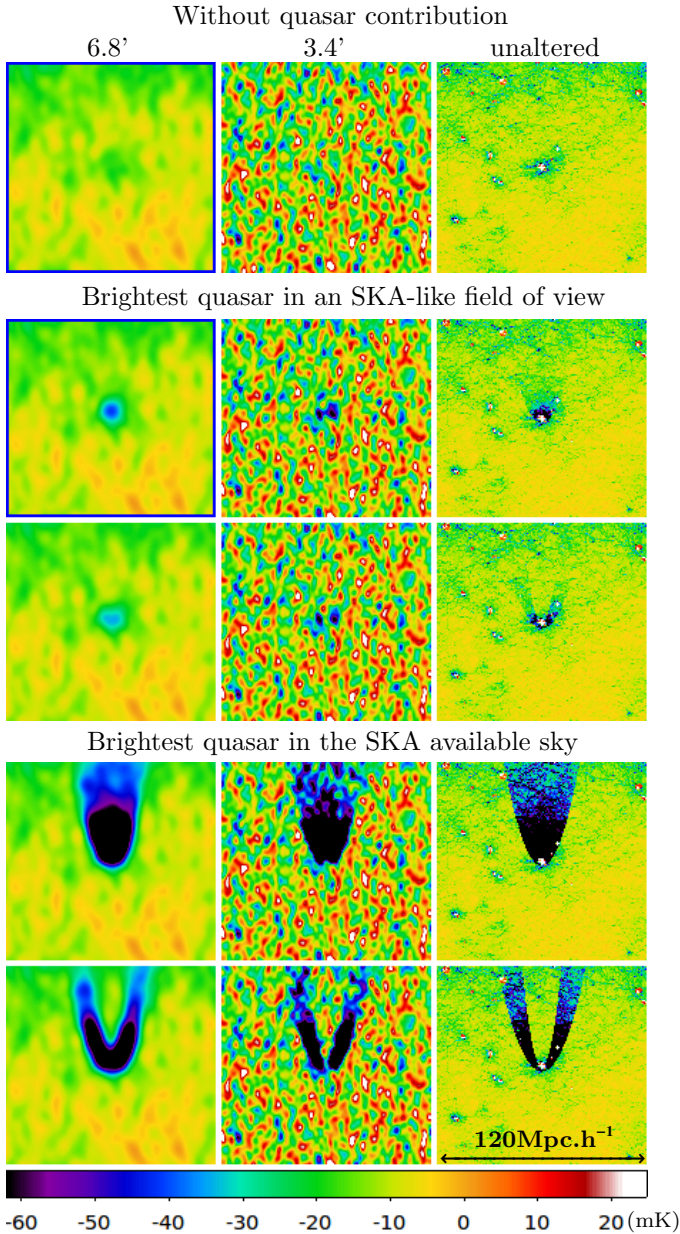

The results are shown in figure \ref{quasarsnoisy}. As can be seen, the brightest quasar in an SKA-like field of view can be seen well above the level of noise in all cases, and the very peculiar shape of the pattern may even be retrievable from the ``timing 2'' $3.4'$ image. Also, we can see the tremendous effect of the brightest quasar of the SKA available sky, with an unmistakable pattern for both timings, at very high S/N ratio. This means that in a standard SKA field, we may be able to put constraints on the length of the duty cycle of one or a few quasars at different redshifts. If a substantial sky survey is made in order to detect a particularly bright quasar, then one could expect tight constraints on the duty cycle. However, one should recall that no prior survey could detect a quasar of the ``timing 2'' type, since it has already turned off when it intersects the lightcone, so a quasar found by any kind of survey would be of the ``timing 1'' type, which only allows us to determine a lower limit on the length of its duty cycle.\\

Finally, one should note that a quasar as bright as the one we expect in the SKA available sky would possibly be found in a different environment. For that reason, we also assigned the same mass to a quasar at $z=9$, in an environment far less advantageous for observability, and we noticed that the S/N ratio, though smaller, is still high, and the pattern clearly identifiable.

\subsection{Collective effect of all active quasars}
In this section, just as before, we assess the combined effect of all the quasars in the field. We first run the FoF halo finder on the different snapshots of the simulation, and recreate a lightcone of haloes. For each halo, we assign a $\SI{1450}{\angstrom}$ luminosity following equation \ref{eq1450}. For simplicity we assume the spectrum to be flat in those regions, the corrective factor between $\SI{1450}{\angstrom}$ and the lyman band (between $\SI{1025}{\angstrom}$ and $\SI{1215}{\angstrom}$) being close to one. Then, we randomly choose a starting time for the duty cycle of our quasar, in the 100\,Myrs before the quasar worldline intersects the past lightcone of the observer. For our fiducial model, with $f_\text{duty}=0.02$, the quasar will shine for 2\,Myr, and for our model with $f_\text{duty}=0.2$, it will shine for 20\,Myr. The length of the duty cycle is somehow an arbitrary choice, and it does change the form of the contribution, but we expect the mean effect not to vary for a different choice of duty cycle length if the same $f_\text{duty}$ is used.\\
\begin{figure*}
\begin{picture}(500,320)
 \centering
 \put(0,0){\includegraphics[width=\textwidth]{allquasarswithout.png}}
 \put(0,100){\includegraphics[width=\textwidth]{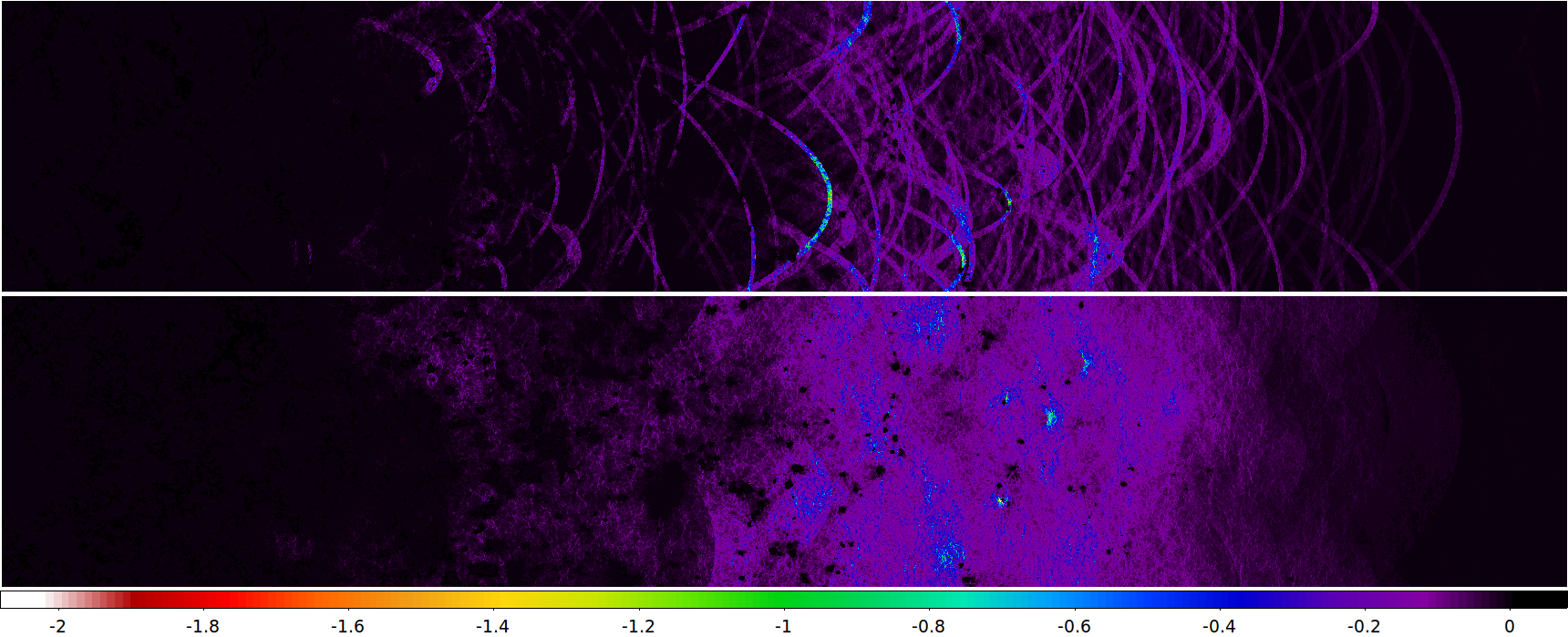}}
\put(498,56){$\xupdownarrow{1.55cm}$}
 \put(488,38){\rotatebox{90}{$200\text{Mpc}.h^{-1}$}}
 \put(414,122){{$\color{white}f_\text{duty}=0.2,~f_\text{corr}=0.2$}}
 \put(417,218){{$\color{white}f_\text{duty}=0.02,~f_\text{corr}=1$}}
 \put(495,1.3){(mK)}
 \put(491,101.3){(mK)}
 \put(0,315){$z$}
 \put(41.5,315){$7$}
 \put(42,306){\bf|}
\put(171.7,315){$8$}
 \put(172.2,306){\bf|}
\put(276.2,315){$9$}
 \put(276.7,306){\bf|}
\put(358.5,315){$10$}
 \put(360.9,306){\bf|}
\put(429.5,315){$11$}
 \put(432,306){\bf|}
\put(489.4,315){$12$}
 \put(492,306){\bf|}
  \end{picture}
 \caption{From top to bottom: Lightcone of $\delta T_b$ (in mK) showing the contribution of the quasars to $\delta T_b$ with $f_\text{duty}=0.02$ and $f_\text{corr}=1$, the contribution of the quasars to $\delta T_b$ with $f_\text{duty}=0.2$ and $f_\text{corr}=0.2$, and the total signal without quasars.}
 \label{allquasarsxa}
\end{figure*}

In figure \ref{allquasarsxa} the lightcone of $\delta T_b$ and the contributions of the quasars for the two models are shown. One can see that the impact of the Lyman band emission is stronger than that of the radio emission, since one can see extended regions with a $\delta T_b$ decreased by at least $\SI{1}{\milli\kelvin}$. However, the effect does not considerably alter the signal. It changes the mean $\delta T_b$ by $\SI{0.2}{\milli\kelvin}$. Although small, this may be detectable in the power spectrum (if sufficiently peaked in terms of characteristic scale).
 \subsection{Power spectrum}
\begin{figure}
 \includegraphics[width=\columnwidth]{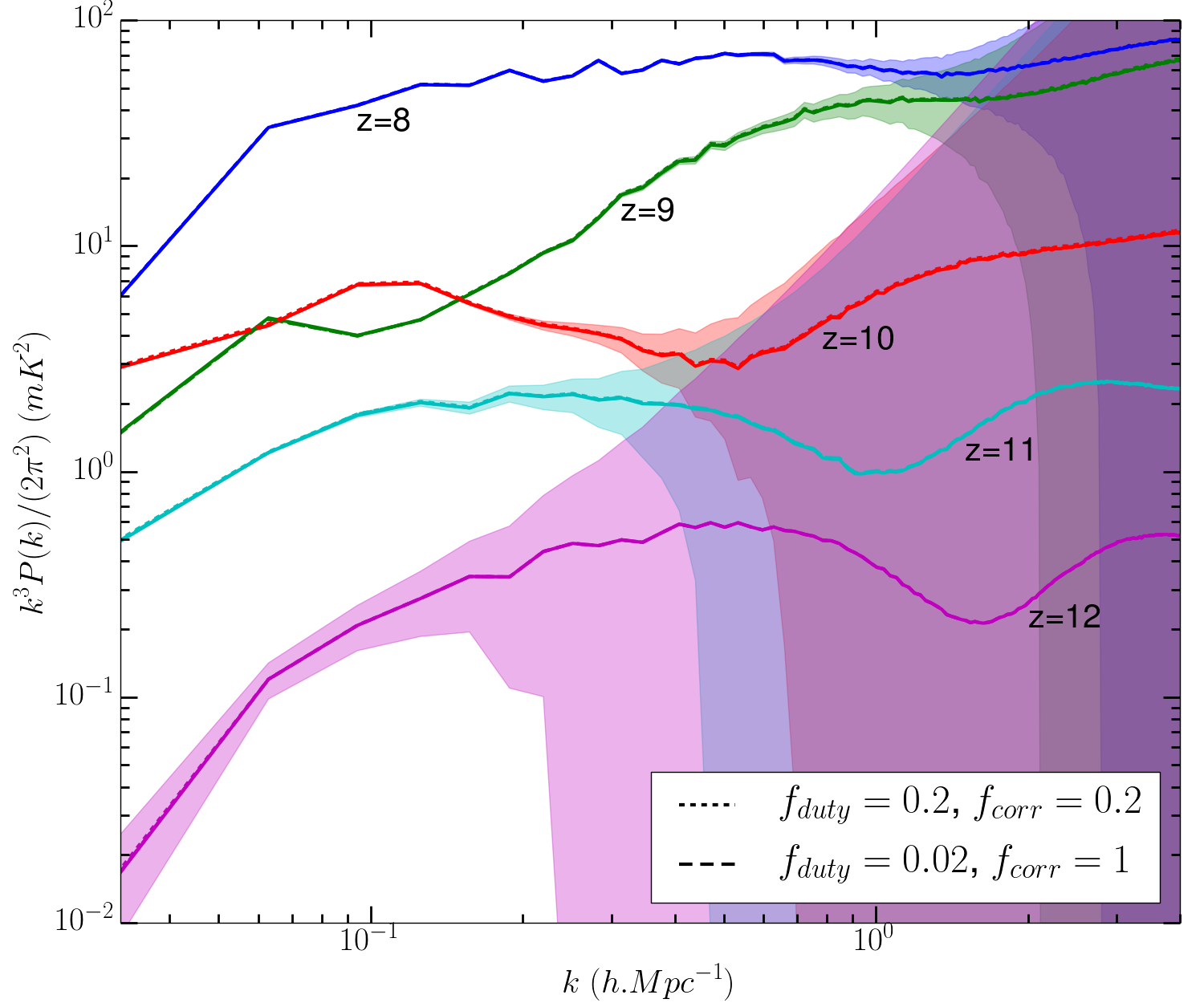}
 \caption{Power spectra for different $z$, and for different setups. The solid lines represent the power spectra without contribution of the quasars, the small dashes our fiducial model with $f_\text{duty}=0.02$ and $f_\text{corr}=1$ and the long dashes the model with $f_\text{duty}=0.2$ and $f_\text{corr}=0.2$. The later two are almost indistinguishable from the first. Shaded areas correspond to the expected error bars of the SKA1, with 1000\,hours of exposure.}
 \label{powerspectraxa}
\end{figure}

\begin{figure}
 \includegraphics[width=\columnwidth]{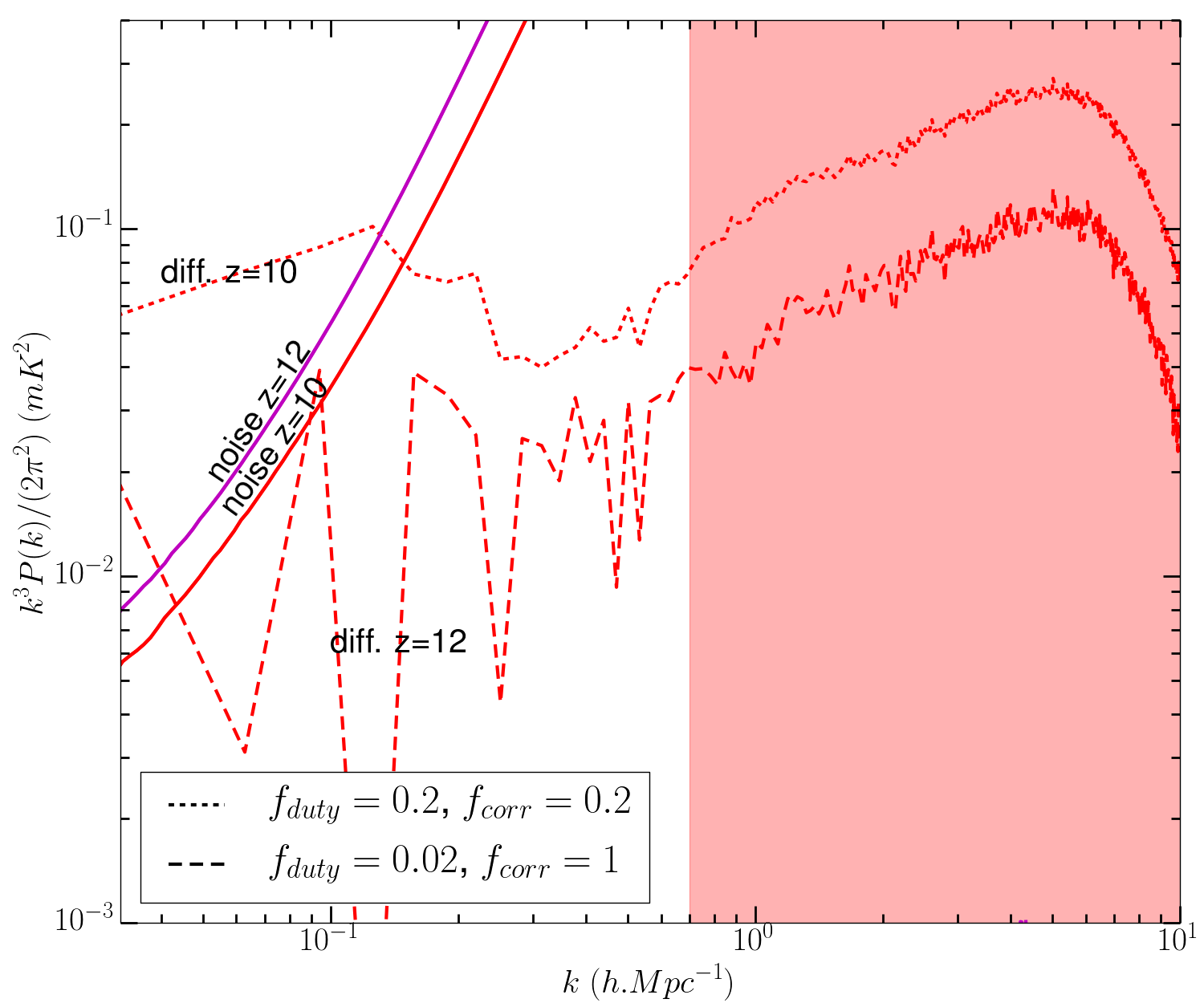}
 \caption{Contribution of the quasars to the power spectrum for $z=10$ and $z=12$, for our two models, compared to the level of noise expected with SKA1, with 1000h of exposure. The shaded area corresponds to the point where the noise for the $z=10$ power spectrum becomes greater than the signal itself.}
 \label{diffpowerspectraxa}
\end{figure}

We computed the power spectra at different redshifts (8, 9, 10, 11 and 12) and for the different models (no quasar, $f_\text{duty}=0.02$ and $f_\text{duty}=0.2$). These can be found in figure \ref{powerspectraxa}, where we also add the error bars corresponding to the thermal noise, in order to assess the detectability of the effect. We notice that the relative effect is mostly invisible, and does not change the shape of the power spectra. However, except for $z=12$, the effect is within SKA sensitivity for low $k$. In order to illustrate this, we plot the differences for two different redshifts ($z=12$ and $z=10$) in figure \ref{diffpowerspectraxa}, as well as the expected level of noise for the SKA. We notice that at $z=10$, the effect is above thermal noise at low $k$, but the difference is featureless, which would make it degenerate with other effects, such as the efficiency of $x_\alpha$ in galaxies. Also, it seems that the difference of the power spectra at $z=10$ exhibits some structure on scales of $\sim\SI{20}{\mega\pc}$ ($k=0.3h\,\text{Mpc}^{-1}$), which are characteristic of the imprint of individual quasars, but the features are unfortunately below the noise level.
\section{Conclusions}
In this work we developped a model for quasar emission in the optical and the radio able to reproduce the observational data available up to $z=6$ (section \ref{sectionmodelquasar}). We then extrapolated this model at redshifts between 6 and 12 and predicted the abundances of radio loud quasars at high redshift. This led us to conclude that, even neglecting the attenuating of synchrotron emission due to the CMB at high redshift, a significant fraction of the sky would have to be surveyed to find a quasar in the $10$\,mJy range, as is usually assumed for 21-cm forest predictions, and that it is unlikely to find such a quasar in a typical SKA field.\\

In section \ref{section21cm} we studied the effect of the radio emission of a quasar on the spin temperature in the surrounding IGM, an effect that has never been taken into account before. Applying this effect to quasars in one cosmological simulation of the 21SSD database (\cite{21SSD}), we concluded that a $\SI{1}{\mJy}$ or $\SI{10}{\mJy}$ quasar may have a substantial effect on the $\delta T_b$ tomography (up to $\SI{40}{\milli\kelvin}$). However, due to the rather small scale of the effect, we showed that it has little chance of being detectable with the SKA. We also concluded that in a typical SKA1-Low field, the combined effect of the radio-loud quasar population is negligible.\\

Finally, in section \ref{sectionxa} we studied the impact of the lyman continuum emission of quasars on the local value of $x_\alpha$ in the IGM, and concluded that the brightest quasar in a typical SKA field will have a substantial impact on the tomography (a difference larger than 40\,mK in regions of several tens of Mpc in size). We also concluded that the specific pattern created by such a quasar may even be detectable in the tomography. Also, the brightest quasar expected in the SKA sky would have an effect so important that the pattern would be clearly identifiable, and one could precisely compute the length of its duty cycle. On the other hand, the overall impact on the tomography is rather modest (up to -1\,mK, and on average -0.2\,mK) and is far below the level of noise. We also studied the impact on the power spectrum, and concluded that it was above the level of noise at low $k$ and $z \leq 11$. However, without any identifyable feature, it will probably be degenerate with other effects such as the efficiency of $x_\alpha$ in galaxies.\\

As already pointed out, for simplicity we assumed direct relations between the halo mass and the black hole mass (for Eddington ratio or for bolometric corrections). These relations are never as tight and if we had included a scatter, it would have decreased the slope of our luminosity functions. In the range of luminosity covered by the observations, every model is more or less equivalent, but a higher slope will mean fewer quasar below the observational range, which means that the luminosities of the quasars in our simulation box (or in a typical SKA field) will be reduced. However, a rarer and more luminous quasar such as the one we expect in the SKA available sky will barely be affected by this change.

The two effects we present in this work are instantaneous effects, and depend strongly on the timing between the duty cycle and the passing of the lightcone. On the contrary, the ionizing emissions (UV and X-ray) of the quasar will mostly act as a cumulative effect. In a forthcoming work we will address this question by running a consistant cosmological simulation, including UV and X-ray emission from both the quasars and the stars. The effects, although less distinctive for single object, may be stronger on the power spectrum.

\section*{Acknowledgements}

This work was made in the framework of the French ANR funded project ORAGE (ANR-14-CE33-0016). We also acknowledge the support of the ILP LABEX (under the reference ANR-10-LABX-63) within the Investissements d'Avenir programme under reference ANR-11-IDEX-0004-02. The simulations were performed on the GENCI national computing center at CCRT and CINES (DARI grants number 2016047376 and A0010407376).




\bibliographystyle{mnras}
\bibliography{/home/bolgar/licorice/Redaction/Biblio,ref_intro} 







\bsp	
\label{lastpage}
\end{document}